\documentclass[12pt,a4paper]{article}
\usepackage{t1enc}
\usepackage[dvips,final]{graphicx} 
\usepackage{tabularx} 
\usepackage{amsfonts}

\def\du{\unskip\smash{\lower 1.4ex \hbox{\char34}}\kern-.2ex}
\def\hu{\kern-.2ex\hbox{\char92}}

\newcommand{\bdis}{\begin{displaymath}}
\newcommand{\edis}{\end{displaymath}}
\newcommand{\be}{\begin{equation}}
\newcommand{\ee}{\end{equation}}

\begin{document}
\baselineskip=6mm
\newpage

\title{A NOTE ON DERIVATION OF RUTHERFORD FORMULA WITHIN BORN APPROXIMATION}
\author{Michal Demetrian\footnote{{\it demetrian@fmph.uniba.sk}} \\
Department of Theoretical Physics \\
Faculty of Mathematics, Physics and Informatics \\
Mlynska Dolina F2, 842 48 Bratislava IV \\
Slovak Republic}
\maketitle

 \abstract{It is shown in this paper that one does not need to use just
 exponential dumping factor when computing the Rutherford formula within Born
 approximation. Text, which is very simple,
 might be of interest for physics students as well as for
 physics teachers.}

 \section*{Example 1}
 The scattering amplitude for a particle of mass $m$ in the spherically
 symmetric potential energy $V(r)$ within the Born approximation is
 given by
 \be \label{sa}
 f(\theta)=-\frac{2m}{\hbar^2\kappa}\int_0^\infty{\rm d}r
 r V(r)\sin(\kappa r) \quad ,
 \ee
 where $\vec{\kappa}=\vec{q}-\vec{q}'$, $\vec{q}$ is the wave vector of
 incident particle and $\vec{q}'$ is the wave vector of scattered
 particle. The vector $\kappa$ is related to the scattering axial angle $\theta$ by
 the equation $\kappa=2q\sin(\theta/2)$. If we insert the Coulomb law
 \bdis
 V(r)=\frac{e^2}{4\pi\epsilon_0 r}
 \edis
 into (\ref{sa}) instead of $V$ we get
 \bdis f(\theta)=-\frac{2m}{\hbar^2\kappa}\frac{e^2}{4\pi\epsilon_0} \int_0^\infty{\rm d}r
 \sin(\kappa r) \quad .
 \edis
 The above written integral does not
 converge, of course. The standard procedure to make the integral
 to have a sense is to regularize the Coulomb law by multiplying it
 by the function $\exp(-\mu r)$, where $\mu>0$ and after having
 performed integration to make the limit $\mu\to 0^+$. We will show
 that the regularizator has not to be of the form $\exp(-\mu r)$ . \\

 Let $\{g_n(r)\}_{n=1}^\infty$ be the sequence of monotonically decreasing
 functions defined on the half-line $r\geq 0$ such that
 $\{g_n\}$ converges point-wise to $1$ and we request that
 $g_n(0)=1$ and $\lim_{r\to\infty}g_n(r)=0$.
 Moreover we request that the derivatives $g'_n(r)$ approach zero
 at $r\to\infty$ uniformly with respect to $n$.  \\
 The scattering amplitude (\ref{sa}) for the potential energy
 \bdis
 V_n(r)=\frac{e^2}{4\pi\epsilon_0}\frac{1}{r}g_n(r)
 \edis
 is given by
 \bdis
 f_n(\theta)=-\frac{2m}{\hbar^2\kappa}\frac{e^2}{4\pi\epsilon_0}\int_0^\infty{\rm d}r
 \sin(\kappa r)g_n(r)\equiv
 -\frac{2m}{\hbar^2\kappa}\frac{e^2}{4\pi\epsilon_0}I_n(\kappa) \quad .
 \edis
 The integral $I_n(\kappa)$ exists (as Riemann integral) due to our assumptions
 on the functions $g_n$ and the Dirichlet's criterion.  Let us compute the integral
 $I_n(\kappa)$ {\it per partes}. We get
 \bdis
 I_n(\kappa)=\frac{1}{\kappa}+\frac{1}{\kappa}\int_0^\infty{\rm d}r\cos(\kappa r)
 g_n'(r) \quad .
 \edis
 Our task is to show that the second term goes to zero as $n$ goes to infinity. But this is
 true because of the properties of the sequence $\{g_n\}$ and because of the fact that there
 exists such positive constant $K>0$ that for all $A>0$ we have
 \bdis
 \left| \int_0^A{\rm d}x\cos(\kappa x)\right|\leq K \quad .
 \edis
 Therefore we have
 \bdis
 \lim_{n\to\infty}f_n(\theta)=-\frac{2m}{\hbar^2}\frac{e^2}{4\pi\epsilon_0}\frac{1}{\kappa^2}
 \quad ,
 \edis
 from which we get the differential cross-section
 \bdis
 \frac{{\rm }d\sigma}{{\rm d}\Omega}=|f_n(\theta)|^2=\frac{4m^2e^4}
 {\hbar^4(4\pi\epsilon_0)^2\kappa^4}
 \edis
 which is nothing else but the Rutherford formula. \\
 As an example we can take the sequence of functions
 \bdis
 g_n(r)=1-\frac{2}{\pi}\arctan\left(\frac{r}{n}\right) \quad .
 \edis
 In this case we are able to compute $f_n(\theta)$ in terms of elementary functions. In fact:
 \begin{eqnarray*}
 I_n(\kappa) & = & \frac{1}{\kappa}-\frac{1}{\kappa}\frac{2}{n\pi}\int_0^\infty{\rm d}r
 \frac{\cos(\kappa r)}{1+\left(\frac{r}{n}\right)^2}=
 \frac{1}{\kappa}-\frac{1}{\kappa}\frac{1}{\pi}\int_{-\infty}^\infty{\rm d}x
 \frac{\exp(i\kappa nx)}{1+x^2} \\
 & = & \frac{1}{\kappa}\left[ 1-2\pi i\lim_{z\to i}(z-i)\frac{\exp(i\kappa n z)}
 {1+z^2}\right]=\frac{1}{\kappa}\left[1-\exp(-\kappa n)\right]\to^{n\to\infty}
 \frac{1}{\kappa} \quad .
 \end{eqnarray*}
 We mention, that it is possible to take less limiting assumption on $g_n$
 to get the same result. \\

 \section*{Example 2}
 In this section we will show another possibility how to
 regularize the Coulomb potential.
Now, let us take the potential
 \be \label{po}
V(r,a)=\frac{e^2}{4\pi\epsilon_0}\frac{1}{r^{1+a}}, \quad 0<a<2
\quad .
 \ee
 Then the scattering amplitude has the form
 \be
 \label{sa1} f(\theta,a)=-\frac{2m}{\hbar^2
 \kappa}\frac{e^2}{4\pi\epsilon_0} \int_0^\infty{\rm d}r
 \frac{\sin(\kappa r)}{r^a} \quad . \ee
 So, we have to compute the
 following integral
 \be \label{int} I(a,\kappa)=
 \int_0^\infty{\rm d}r\frac{\sin(\kappa r)}{r^a} \quad . \ee
 To do this it would be
useful to realise that the following identity holds \be \label{id}
\frac{1}{x^a}=\frac{1}{\Gamma(a)}\int_0^\infty{\rm d}t
t^{a-1}e^{-tx} \quad . \ee Inserting (\ref{id}) into (\ref{int})
we have
\begin{eqnarray*}
I(a,\kappa) & = & \frac{1}{\Gamma(a)}\int_0^\infty{\rm d}t t^{a-1}
\int_0^\infty{\rm d}r\sin(\kappa r)e^{-tr}=\frac{\kappa}{\Gamma(a)}
\int_0^\infty{\rm d}t \frac{t^{a-1}}{\kappa^2+t^2} \\
& = & \left| t=\kappa u\right| =
\frac{\kappa^{a-1}}{\Gamma(a)}\int_0^\infty
{\rm d}u\frac{u^{a-1}}{1+u^2}=\left| \frac{1}{1+u^2}=w\right| \\
& = & \frac{\kappa^{a-1}}{2\Gamma(a)}\int_0^1{\rm d}w
(1-w)^{a/2-1}w^{a/2}=\frac{\kappa^{a-1}}{2\Gamma(a)}
\frac{\pi}{\sin\left(\frac{\pi a}{2}\right)} \quad .
\end{eqnarray*}
So, the scattering amplitude for the potential (\ref{po}) is given by
the following formula
\be \label{sa2}
f(\theta,a)=-\frac{2me^2}{4\pi\epsilon_0\hbar^2\kappa^{2-a}}
\frac{\pi}{2\Gamma(a)\sin\left(\frac{\pi a}{2}\right)} \quad ,
\ee
from which we get in the limit $a\to 0^+$ the amplitude
\bdis
f(\theta)=-\frac{2me^2}{4\pi\epsilon_0\hbar^2\kappa^2}
\edis
which leads just to the Rutherford formula. \\

\end{document}